# CONTROLLING THE ĆUK CONVERTER USING PIECEWISE LINEAR LYAPUNOV FUNCTIONS

**Aleksandra Lekić, Dušan Stipanović\*, Nikola Petrović**

School of Electrical Engineering, University of Belgrade, Belgrade, Serbia
\*Coordinated Science Laboratory, University of Illinois at Urbana-Champaign, Urbana, USA

**Abstract:** *In this paper we design a switching control law for the Ćuk converter in the continuous conduction mode using piecewise linear Lyapunov functions. These Lyapunov functions can be constructed using different number of state variables affecting the system's performance. In the paper, some representative simulations covering construction of different piecewise Lyapunov functions, are provided.*

## 1. INTRODUCTION

Designing stabilizing controllers for DC-DC converters has been a very popular and demanding research topic [3] over a number of years. In general, there are many approaches designed to guarantee converters' outputs satisfying the prescribed specifications. Most popular approaches are based on first performing the averaging of the converters' variables and constructing PID controllers to meet the desired performance specifications [4], [9]. Nowadays, the PID control of DC-DC converters is implemented digitally using digital signal processors or using FPGAs as described in [9]. Another well-known control approach is the current mode control approach used in a few variations where the most popular are hysteresis window and peak limiting methods [4], [10]. In particular, the peak limiting control approach should be carefully designed because if done otherwise it may easily result in the controlled converter's chaotic [10].

In general, DC-DC converters are circuits consisting of switches that are reactive components. Depending on the positions of the switches, each DC-DC converter can operate in a few operating modes that are behaving as linear-time invariant dynamic systems. However, the switching among these modes implies that the overall behaviors of DC-DC converters have to be treated as nonlinear. Thus, a lot of research effort has been concentrating on developing nonlinear control design methods. Comparing to the classical PID control, the nonlinear control design methods produces smaller dynamic losses [2].

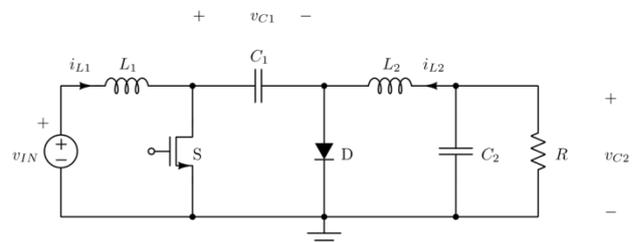

Fig. 1. *Ćuk converter.*

Since we want not only to stabilize the switched system but also to control the ripple, a control law based on the switching among multiple piecewise linear Lyapunov functions [1], [8] having polytopic level sets, is designed in this paper. The earlier results reported in [1], [8] provide different algorithms for the fast construction and computation of the level set polytopes thus proving fast convergence. There are some other approaches for constructed polytopic functions like ray gridding algorithm [12], [13] which is successfully applied for control of the Boost converter considering the control of its averaged model. However, there are approaches that can be applied to all switched systems which can be stable, unstable or marginally stable [7]. This approach will be used for checking the obtained results in this paper.

In this paper, we consider the Ćuk converter as depicted in Fig. 1 which operates in the continuous conduction mode (CCM). The continuous conduction mode of the Ćuk converters turns out to be very specific because the converter switches between two marginally stable subsystems. This marginal stability property implies that we need two piecewise linear Lyapunov functions (PLLF) with specific polytopic value sets. It is also shown that the Ćuk converter operating in the continuous conduction mode forms a system which can be decoupled with respect to the two specific dynamics: first being the dynamics of inductor current $i_{L1}$ and capacitor voltage $v_{C1}$; and the other one being the dynamics of induct or current $i_{L2}$ and capacitor voltage $v_{C2}$. Inductor current $i_{L1}$ is an unstable state variable and for that matter it has to be particularly used in the polytopic level set construction. Combined dynamics of inductor current $i_{L1}$ and capacitor voltage $v_{C1}$ is used to determine the converter's steady state and the ripple of the state variables in the steady state. However, it can be shown that the dynamics of inductor current $i_{L2}$ and capacitor voltage $v_{C2}$ is used only for the converter's initial transient for the faster transition to the steady-state and to obtain a smaller overshoot.

The paper is organized as follows. In Section II, the models of the Ćuk converter while operating in CCM operation as well as the constraints on the construction of the piecewise linear Lyapunov functions, are provided. In Section III, simulation results for four constructed polytopes are given. The paper is concluded with some remarks in Section IV.

## 2. CONTROL OF THE ĆUK CONVERTER

### 2.1. Operation of the Ćuk converter

Ćuk converter is one of the most complex fourth order DC-DC converters. It is constructed using two switches being transistor S and a diode D. The switchings produce four operating subsystems and two of them are operating in the CCM. First operating subsystem occurs when the PWM signal on the gate of transistor S is on the logical "1" and then with the notation $\mathbf{x} = [i_{L1}\ i_{L2}\ v_{C1}\ v_{C2}]^T$ the system is described with the following state-space equation in the matrix form:

$$\dot{\mathbf{x}} = \begin{bmatrix} 0 & 0 & 0 & 0 \\ 0 & 0 & \frac{1}{L_2} & \frac{1}{L_2} \\ 0 & -\frac{1}{C_1} & 0 & 0 \\ 0 & -\frac{1}{C_2} & 0 & -\frac{1}{RC_2} \end{bmatrix} \mathbf{x} + \begin{bmatrix} \frac{v_{IN}}{L_1} \\ 0 \\ 0 \\ 0 \end{bmatrix} = \mathbf{A}_1 \mathbf{x} + \mathbf{B}_1. \quad (1)$$

The second operating subsystem that occurs during operation in CCM is the mode when the transistor S is turned off and the diode conducts. Then the system is described with equation in matrix form

$$\dot{\mathbf{x}} = \begin{bmatrix} 0 & 0 & -\frac{1}{L_1} & 0 \\ 0 & 0 & 0 & \frac{1}{L_2} \\ \frac{1}{C_1} & 0 & 0 & 0 \\ 0 & -\frac{1}{C_2} & 0 & -\frac{1}{RC_2} \end{bmatrix} \mathbf{x} + \begin{bmatrix} \frac{v_{IN}}{L_1} \\ 0 \\ 0 \\ 0 \end{bmatrix} = \mathbf{A}_2 \mathbf{x} + \mathbf{B}_2. \quad (2)$$

Operation in the CCM is obtained with a switching between the two subsystems during one switching period $T_S$. During the time interval $0 \leq t < dT_S$ converter operates in subsystem 1, in which the transistor S conducts and the diode is turned off (here is assumed that the initial time is 0 and $d$ is duty-ratio of the PWM signal). Afterwards, during the time interval $dT_S \leq t < T_S$ converter operates in subsystem 2 described with the system of equations (2).

If we apply Volt-second and Ampere-second balance in order to determine converter's equilibrium [4], we obtain

$$\bar{\mathbf{x}} = \begin{bmatrix} \bar{\imath}_{L1} \\ \bar{\imath}_{L2} \\ \bar{v}_{C1} \\ \bar{v}_{C2} \end{bmatrix} = \begin{bmatrix} \frac{d^2 v_{IN}}{(1-d)^2 R} \\ \frac{d v_{IN}}{(1-d)R} \\ \frac{v_{IN}}{1-d} \\ -\frac{d v_{IN}}{1-d} \end{bmatrix}. \quad (3)$$

Absolute change of the state variables' values in the steady-state can be calculated then as:

$$\begin{aligned} \Delta x_1 &= \frac{v_{IN} d T_S}{L_1}, \\ \Delta x_2 &= \frac{v_{IN} d T_S}{L_2}, \\ \Delta x_3 &= \frac{v_{IN} d^2 T_S}{(1-d) R C_1}, \\ \Delta x_4 &= \frac{d v_{IN} T_S^2}{8 L_2 C_2}. \end{aligned} \quad (4)$$

Equations (3) and (4) will be used for the control construction. In the further text will be used variables $\mathbf{y} = \mathbf{x} - \bar{\mathbf{x}}$ and their corresponding differential equations are of the form $\dot{\mathbf{y}} = \mathbf{A}_i \mathbf{y} + \bar{\mathbf{B}}_i$ for both subsystems $i = 1,2$. According to equations (1) and (2) the corresponding matrices $\bar{\mathbf{B}}_i$ are

$$\bar{\mathbf{B}}_1 = \begin{bmatrix} \frac{v_{IN}}{L_1} & \frac{v_{IN}}{L_2} & -\frac{d v_{IN}}{(1-d) R C_1} & 0 \end{bmatrix}^T, \quad (5)$$

and $\bar{\mathbf{B}}_2 = a \bar{\mathbf{B}}_1$, where $a = -\frac{d}{1-d}$.

### 2.2. Construction of piecewise linear Lyapunov functions

Piecewise linear Lyapunov functions are defined as

$$V(\mathbf{y}) = \max_{j \in \mathbf{J}} |k_j y_j|, \quad (6)$$

where $\mathbf{J} = \{1,2,3,4\}$, $k_j$ are corresponding state variables' coefficients and $\mathbf{y} = \mathbf{x} - \bar{\mathbf{x}}$. Defined Lyapunov functions (6) are invariant on the attractive set or polytope $\mathbf{P}$ given by

$$\mathbf{P} = \{\mathbf{y} \in \mathbb{R}^4 : |k_j y_j| \leq 1, j \in \mathbf{J}\}, \quad (7)$$

with the stabilizing switching rule will be as follows:

$$q = \begin{cases} 0 & \text{if 1 and } k_j y_j \geq 1, \\ \{0,1\} & \text{if } \{0,1\} \text{ and } -1 < k_j y_j < 1, \quad j \in \mathbf{J}. \\ 1 & \text{if 0 and } k_j y_j \leq -1, \end{cases} \quad (8)$$

Polytope $\mathbf{P}$ is attractive set which can be proved by determining the sign of the Lyapunov function derivative on the polytope facets [6]. Polytope $\mathbf{P}$ can be constructed using all circuit state variables, but it can be constructed with the less variables as well.

For the desired duty-ratio and switching period and by knowing the circuit parameters, coefficients $k_j, j \in \mathbf{J}$, can be found [6] as:

$$\begin{aligned} k_1 &= \frac{L_1}{\rho}, \quad |k_2| < \frac{L_2}{\rho}, \\ k_3 &= -\frac{(1-d) R C_1}{d \rho}, \\ |k_4| &< \frac{8 L_2 C_2}{\rho T_S}, \end{aligned} \quad (9)$$

where $\rho = \frac{d v_{IN} T_S}{2}$. Previous equation (9) provides way to construct PLLF control with exactly specified equilibrium and ripple in the steady-state as given in (4). Ćuk converter operating in CCM is stable in the respect to constructed polytope $\mathbf{P}$. For the design of the PLLF control is necessary only to calculate coefficients (9) and to apply the previous switching law (8), which is a benefit comparing to hysteresis control in [5].

In order to provide proof for the coefficients (9) sign, we will use the approach provided in the research [7]. According to the mentioned paper [7], system can be stabilized on the polytope $\mathbf{P}$ if we can construct matrices

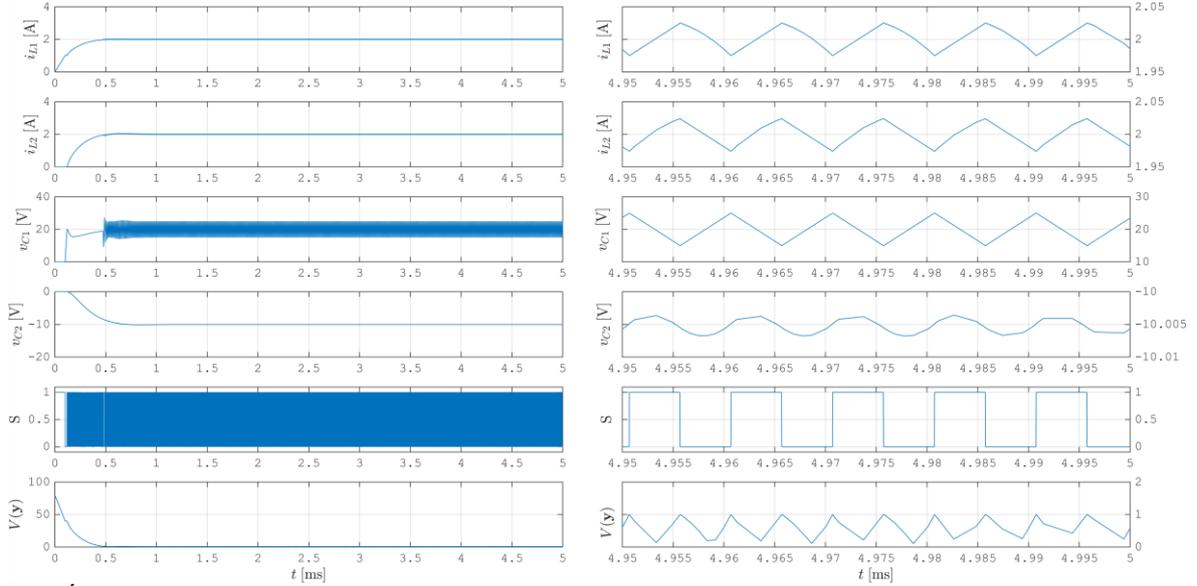

Fig. 2. Ćuk converters' time diagrams, with applied control using PLLF with controlling variables: $i_{L1}$ and $i_{L2}$: left - during first 5ms after circuit's startup and right - during the time interval 4.95ms - 5ms.

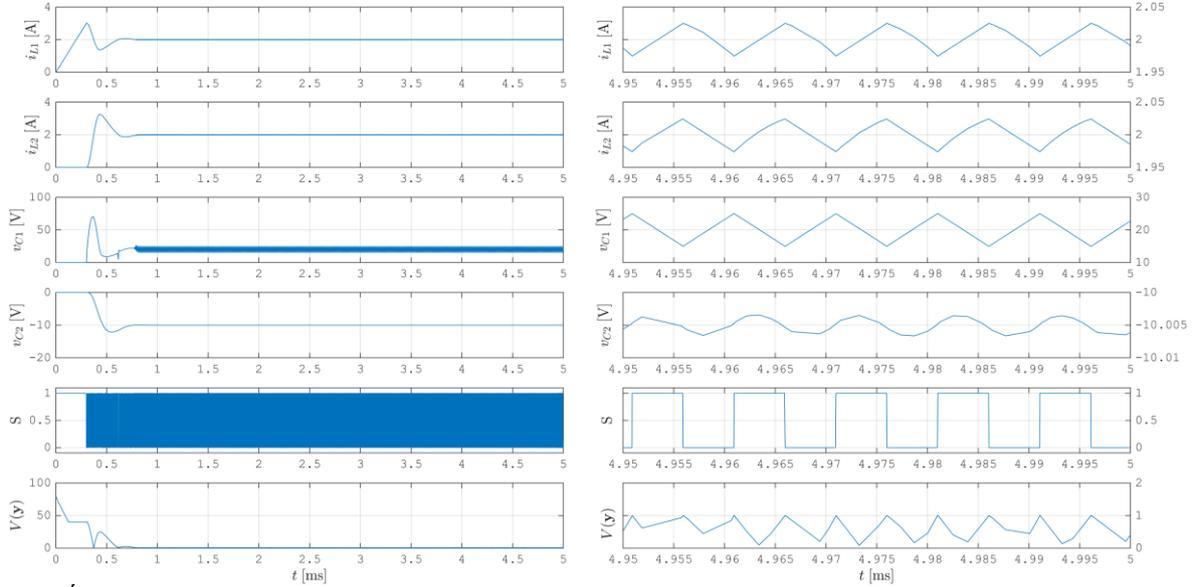

Fig. 3. Ćuk converters' time diagrams, with applied control using PLLF with controlling variables: $i_{L1}$ and $i_{L2}$: left - during first 5ms after circuit's startup and right - during the time interval 4.95ms - 5ms.

$L_i$ and $R_i$ for all state variables $j \in \{1,2,3,4\}$ and for both subsystems $i \in \{1,2\}$ such that

$$L_j A_i R_j < 0 \quad (10)$$

and $L_j R_j = c, c > 0$. In addition to the inequality (10), in the inequality can stand sign "≤" if the system matrix is marginally stable for some state variable.

In order to obtain stable system, we will check if the previous inequality (10) is satisfied for all state variables. Let us first check the variable $i_{L1}$, or $y_1$. It should be $L_1 A_i R_1 < 0$, so if we choose $L_1 = [k_1\ 0\ 0\ 0]$ and $R_1 = [r/k_1\ 0\ r\ 0]^T$, where $L_1 R_1 = r > 0$, we get $L_1 A_1 R_1 = 0$ and $L_1 A_2 R_1 = -\frac{k_1 r}{L_1}$. This is satisfied for $r, k_1 > 0$ which corresponds to equation (9). For the variable $y_2$ we have dependence of the polyhedron $L_2 = [0\ k_2\ 0\ 0]$, so it can be chosen $R_2 = [0\ r\ -r\ -r]^T$ for which is $L_2 R_2 = k_2 r > 0$. By applying the calculation $L_2 A_1 R_2 = -\frac{2 r k_2}{L_2}$ and $L_2 A_2 R_2 = -\frac{r k_2}{L_2}$, so it is again $k_2 r > 0$. Clearly nothing about the sign of $k_2$ cannot be concluded, which gives the possibility to be chosen either positive either negative. Similarly, for variable $y_3$ we can choose $L_3 = [0\ 0\ k_3\ 0]$ and $R_3 = [r\ -r\ r/k_3\ 0]^T$, $L_3 R_3 = r > 0$ and get $L_3 A_1 R_3 = L_3 A_2 R_3 = \frac{k_3 r}{C_1}$, which is satisfied for $k_3 < 0$ as in equation (9). Polytope $P$ depends on variable $y_4$ with the matrix $L_4 = [0\ 0\ 0\ k_4]$ for which can be chosen matrix $R_4 = [0\ r\ 0\ r]^T$, such that $L_4 R_4 = r k_4 > 0$. Now, it is $L_4 A_i R_4 = -r k_4 \left(\frac{1}{C_2} + \frac{1}{R C_2}\right) < 0$, which provides again as a condition $r k_4 > 0$. Clearly, $k_4$ can be both positive or negative, and $r$ is then chosen to satisfy the previous condition $r k_4 > 0$. That means that all state variables have signs as specified by equation (9) and their values are determined by desired steady-state operation. It should be noted more that variable $y_1$ is unstable in both subsystems 1 and 2 and therefore this variable has always to be taken into account while forming the polytope **P**.

### 3. SIMULATION RESULTS

In order to control the converter, we apply the piecewise linear Lyapunov functions control design approach with the following parameters: $L_1 = L_2 = 1\text{ mH}$, $C_1 = 1\text{ μF}$, $C_2 = 20\text{ μF}$, $R = 5\text{ Ω}$ and $v_{IN} = 10\text{ V}$. In this section we will focus on constructing polytope **P** with different number of state variables, so that the set $\mathbf{J} \subset \{1,2,3,4\}$. This control produces Lyapunov-like functions and it may be linked to the partial stability concept as in [11]. Furthermore, this modifications of the set **J** will influence the converter's behavior which will be illustrated by a number of representative simulations.

First we will focus on the polytope constructed only using two state variables - inductor currents $i_{L1}$ and $i_{L2}$, so then the set $\mathbf{J} = \{1,2\}$. As mentioned before, in order to obtain stable system with respect to the desired equilibrium and switching period it is necessary to construct polytope **P** dependent on variable $i_{L1}$, because inductor current $i_{L1}$ presents unstable variable. In the case when $d = 0.5$ and $T_S = 10\text{ μs}$ the resulting equilibrium $\bar{\mathbf{x}} = [2\text{A} \ 2\text{A} \ 20\text{V} \ -10\text{V}]^T$ and calculated coefficients are: $\rho = \frac{dT_S v_{IN}}{2} = 2.5 \cdot 10^{-5}$, $k_1 = \frac{L_1}{\rho} = 40\text{ A}^{-1}$ and $|k_2| = \frac{L_2}{2\rho} = 20\text{ A}^{-1}$. In Fig. 2 we depict the example of the Ćuk converter controlled using PLLF control design where the controlling variables are $i_{L1}$ and $i_{L2}$ with coefficient $k_2 < 0$. In Fig. 3 however the same control is applied, but the coefficient $k_2$ is taken positive. In both Figs. 2 and 3, an initial transient behavior of the Ćuk converter during 5 ms, are depicted left, while on the right portions we provide diagrams from the left part of Figs. 2 and 3 during the interval of time $4.95\text{ ms} - 5\text{ ms}$, in order to zoom out the converter's steady-state operation. In both Figs. 2 and 3, we provide both diagrams of the state variables and the positions of switch S in the following way: "1" when the switch conducts and "0" when the switch which is turned off. It can be also observed that the values of the switched Lyapunov function $V(\mathbf{y})$ is clearly less than 1 during the operation which is one of the design specifications. One of the main benefits of using the piecewise linear or polytopic Lyapunov control design was not only in stabilizing the controller but also in having a very precise estimates of the ripple. From the diagrams in Figs. 2 and 3 is visible that in the case when a negative $k_2$ is chosen, there is no overshoot on state variables initial transient. However, in both cases the steady-state is the same, but $k_2 < 0$ is better for avoid circuit's malfunctioning caused by very high values of some state variables.

In order to compare results when we apply the polytope constructed using two state variables, we have constructed polytopes using three and four variables. In the case of a three variable polytope, we chose variables $i_{L1}$, $i_{L2}$ and $v_{C1}$ with coefficients $\rho = \frac{dT_S v_{IN}}{2} = 2.5 \cdot 10^{-5}$ and $k_1 = \frac{L_1}{\rho} = 40\text{ A}^{-1}$, $k_2 = -\frac{L_2}{2\rho} = -20\text{ A}^{-1}$ and $k_3 = -\frac{(1-d)RC_1}{d\rho} = -0.2\text{ V}^{-1}$. Again, in Fig. 4 are depicted simulation results during first 5 ms left and its portion during time interval $4.95\text{ ms} - 5\text{ ms}$. Comparing to Fig. 3 in this case there is no overshoot on state variables because the coefficient $k_2 < 0$. But, there is advantage over using two state variables as in Fig. 2 which is clearly visible on the diagram for $v_{C1}$. That is, in Fig. 2 there is small overshoot in the ripple for capacitor $C_1$'s voltage which can be seen during initial transient around 0.5 ms. However, the same thing does not occur in the case when we construct three variable polytope as in Fig. 4.

Finally, in Fig. 5, the time diagrams of the Ćuk converter controlled using PLLF for which the polytope is obtained using all four state variables with coefficients $\rho = \frac{dT_S v_{IN}}{2} = 2.5 \cdot 10^{-5}$ and $k_1 = \frac{L_1}{\rho} = 40\text{ A}^{-1}$, $k_2 = -\frac{3L_2}{4\rho} = -30\text{ A}^{-1}$, $k_3 = -\frac{(1-d)RC_1}{d\rho} = -0.2\text{ V}^{-1}$ and $k_4 = -\frac{L_2 C_2}{100\rho T_S} = -0.8\text{ V}^{-1}$, are depicted. In this case it is clear that, again, there is no overshoot. The initial transient is longer which is caused by the inclusion of one more state variable $v_{C2}$.

### 4. CONCLUSION

In this paper we provide a construction of the piecewise linear Lyapunov functions to be used for control of the Ćuk converter. The procedure for constructing these functions with respect to the desired equilibrium and ripple values is given in detail. It is shown how the initial circuit transient behavior changes when the Lyapunov functions are constructed based on different numbers of state variables and this result is complemented with the illustrative numerical simulations.

### 5. ACKNOWLEDGEMENT

This work is supported by project TR33020 of the Ministry of Education, Science and Technological Development of the Republic of Serbia.

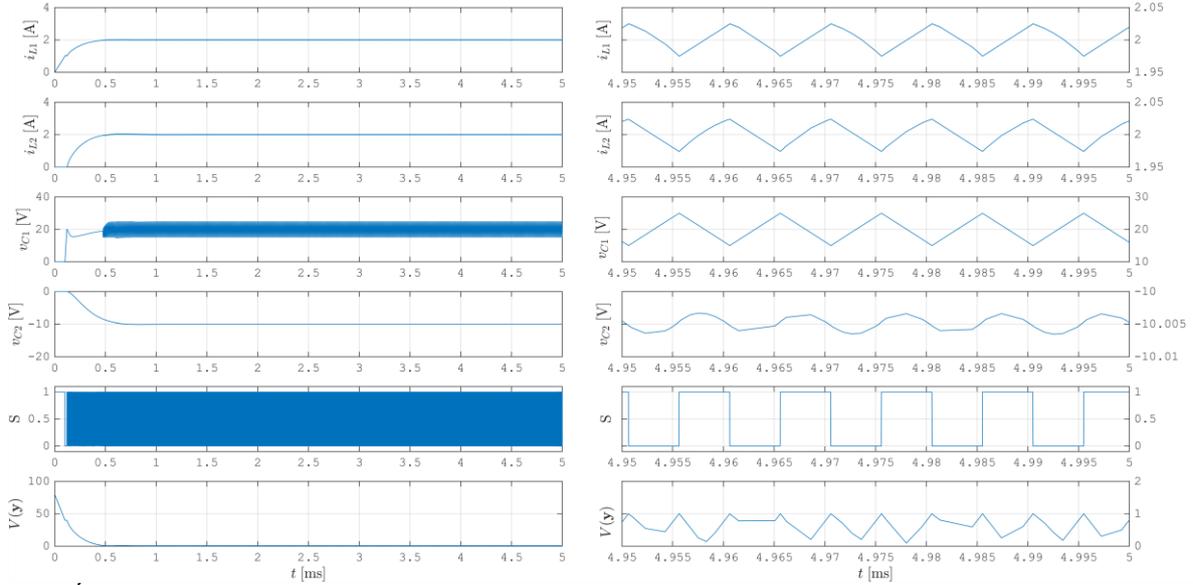

Fig. 4. Ćuk converters' time diagrams, with applied control using PLLF with controlling variables $i_{L1}$, $i_{L2}$ and $v_{C1}$: left - during first 5ms after circuit's startup and right - during the time interval 4.95ms - 5ms.

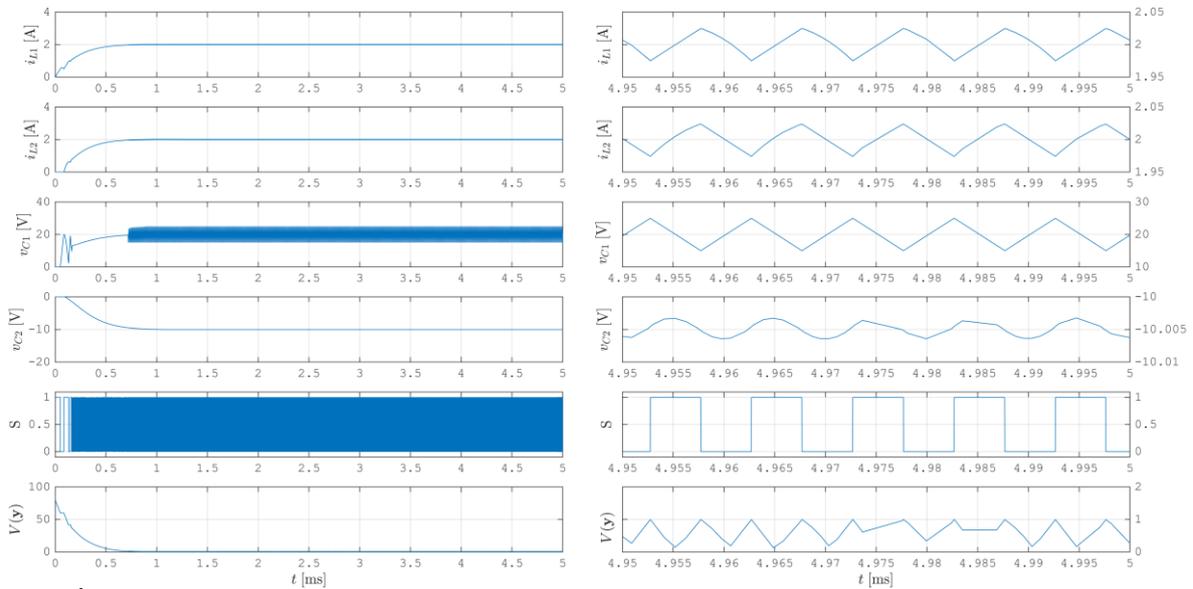

Fig. 5. Ćuk converters' time diagrams, with applied control using PLLF with controlling variables $i_{L1}$, $i_{L2}$, $v_{C1}$ and $v_{C2}$: left - during first 5ms after circuit's startup and right - during the time interval 4.95ms - 5ms.